\documentclass[12pt]{iopart}
\usepackage{graphicx}
\usepackage{subfigure}

\begin{document}

\title[Magnetic phase diagram of NaNiO$_2$]{Magnetic phase diagram of the S=1/2 triangular layered compound NaNiO$_2$: a single crystal study}

\author{S  de Brion$^1$, M Bonda$^2$, C Darie$^1$, P Bordet$^1$ and I Sheikin$^2$}

\address{1 Institut N\'{e}el, CNRS and Universit\'{e} Joseph Fourier, BP166, F-38042 Grenoble cedex 9,
France}

\ead{sophie.debrion@grenoble.cnrs.fr}

\address{2 Grenoble High Magnetic Field Laboratory, CNRS,
BP166, F-38042 Grenoble cedex 9, France}

\begin{abstract}
Using magnetic torque measurement on a NaNiO$_2$ single crystal,
we have established the magnetic phase diagram of this triangular
compound. It presents 5 different phases depending on the
temperature (4 K - 300 K) and magnetic field (0 - 22 T) revealing
several spin reorientations coupled to different magnetic
anisotropies.
\end{abstract}

\pacs{71.27+a, 75.30.Gw, 73.30Kz}

\maketitle

\section{Introduction}
In magnetic systems with a two dimensional triangular lattice,
magnetic frustration may occur depending on the nature of the
magnetic interactions and the orbitals involved. The case of
LiNiO$_2$, where Ni$^{3+}$  has half filled $e_g$ orbitals, has
been studied for a long time both experimentally and theoretically
\cite{BRI05}. Recent progress on the understanding of the orbital
and magnetic properties of this compound has been achieved thanks
to an appropriate control of the non stoichiometry or doping
effects since no pure samples are available. Indeed magnetic
Ni$^{2+}$ ions are substituted between the triangular planes on
the Li sites creating new magnetic exchange paths. It has been
shown that Li$_{1-x}$Ni$_{1+x}$O$_2$ with x=0.01 does not show
long range magnetic order. Rather, a fluctuating state develops
which is slowed down when Mg doping is introduced and no
interplane Ni$^{2+}$ are presents \cite{BOND08}. When Li is
deintercalated, in Li$_{z}$NiO$_2$,  short range incommensurate
antiferromagnetic order is observed for z=3/4 and 2/3
\cite{SUG08}. As for the orbital part, several experimental works
are in favor of the $\vert 3z^{2}-r^{2}\rangle$ orbital occupancy
for the $e_g$ electrons with finite range order \cite{CHU05} or
dynamical effects \cite{BOND08}. Such an occupancy is indeed
observed unambiguously in the parent compound NaNiO$_2$, which
seems to behave more classically. It orders ferro orbitally below
$\sim480$~K \cite{CHA00b} and antiferro magnetically below $20$~K
\cite{BON66}. We present here the first single crystal study of
its magnetic phase diagram up to $20$~T using torque measurements.
We show that several spin reorientations occur as a function of
temperature or magnetic field with unusual magnetic anisotropy.

\section{State of the art on NaNiO$_2$}

\begin{figure}
  \includegraphics [width=8.5cm]{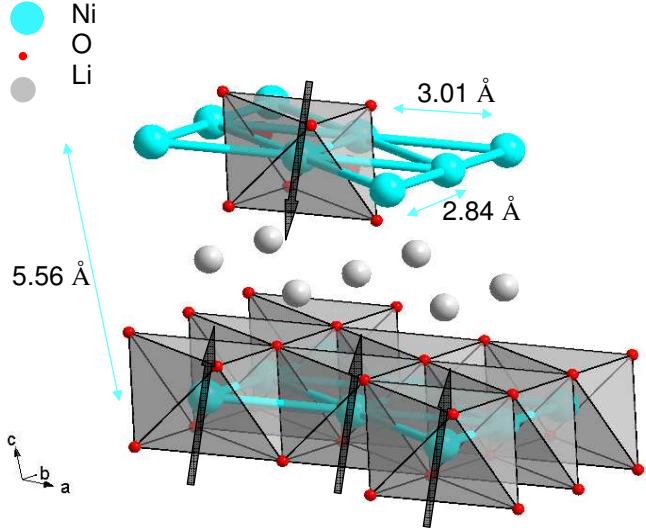}
  \caption{The triangular layered structure of NaNiO$_2$. The arrows indicate the spin directions in the zero field magnetic phase.}\label{Structure}
\end{figure}

 NaNiO$_2$ crystallises in the R-3m  space group and, below
the ferro orbital ordering temperature, becomes monoclinic (C2/m)
with $a$=5.3087~\AA, $b$=2.8413~\AA, $c$=5.5670~\AA, and $\beta=
~110.45^{\circ}$ \cite{DAR05}. The two dimensional triangular
network of Ni$^{3+}$  becomes then distorted to accommodate the
collective elongation of the oxygen octahedra surrounding the Ni
ions (figure \ref{Structure}). At room temperature, the distorted
triangles are described by two angles at 56.3$^{\circ}$ and
61.8$^{\circ}$ and  two Ni - Ni distances at
 2.84~\AA  and 3.01~\AA. From one triangular plane to the other,
 the Ni-Ni distance is 5.47~\AA. The antiferromagnetic ground
 state observed below $T_{\rm N}$ is of the A type with  a ferromagnetic
 alinement of the Ni spins in the triangular plane and an
 antiferromagnetic alinement from one plane to the other.
  The magnetic moments point at 80$^{\circ}$ from the
triangular planes. The Curie Weiss behavior of the magnetic
 susceptibility above 100K leads to ferromagnetic dominant
 interactions with a Curie Weiss temperature $\theta$=+36~K.
 Below the ordering temperature ($T_{\rm N}$=20 K), application of a magnetic field reveals a spin flop
 transition at $\mu_0H_{c1}$=1.8~T \cite{BON66} forT=4~K and another transition at
 $\mu_0H_{E}$=8~T \cite{BAK05,BRI07}. Spin wave measurements agree with an easy plane A type
 antiferromagnet model although there are some discrepancies between the
 inelastic neutron measurements \cite{LEW05} and the electron spin resonance
 measurements \cite{BRI07}. In this model, $H_{E}$ describes the strength of the
antiferromagnetic coupling between the Ni planes. A small
anisotropy in the easy plane is then responsible for the spin flop
transition at $H_{c1}$.

\section{Synthesis of NaNiO$_2$ single crystals}

\begin{figure}
  \includegraphics[width=8.5cm]{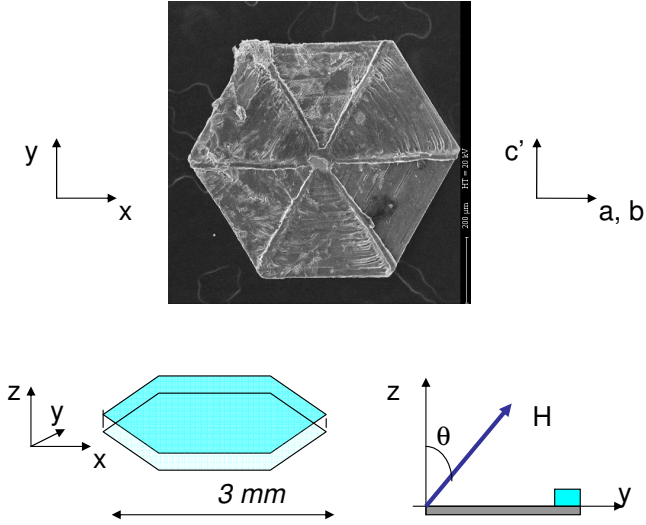}
  \caption{characteristic hexagonal shape of a 0.5$~mm$ NaNiO$_2$ crystal and experimental configuration.}\label{crystal}
\end{figure}

NaNiO$_2$ single crystals have been grown by molten hydroxide flux
method. The flux composition was NaOH and KOH, a Ni crucible
played the role of Ni source. In a preliminary step, 61.6$~g$ of a
NaOH-KOH mixture (molar ratio 1:2) was melted in a separate Ni
crucible at 400$^{\circ}$C. The molten mixture was then poured in
another 40$~mL$ Ni crucible and placed in a preheated open
vertical furnace. The experiment was carried out at
650$^{\circ}$C. The evaporation of the NaOH-KOH mixture at the
liquid surface induces a local saturation, then nucleation and
crystal growth. Runs were stopped once an abundant "crust" was
observed covering the flux surface (typically after 12 hours). The
crystals were easily separated from the flux with water in an
ultrasonic bath. They often take the shape of hexagonal small
platelets (figure \ref{crystal}). The crystals were characterized
using scanning electron microscopy equipped with EDX analysis
(relative content: Na: 0.49 , Ni: 0.51). A mixture of crystals
were crushed and studied by powder X-Ray diffraction: only
stoichiometric NaNiO$_2$ was identified. Most of the crystals are
twinned in the $a, b$ plane so that the crystal platelets contain
three kinds of domains with an average $c^{'}$ direction
perpendicular to the platelets plane which coincides with the Ni
planes.

\section{Magnetic torque measurements}

\begin{figure}
  \includegraphics[width=8.5cm]{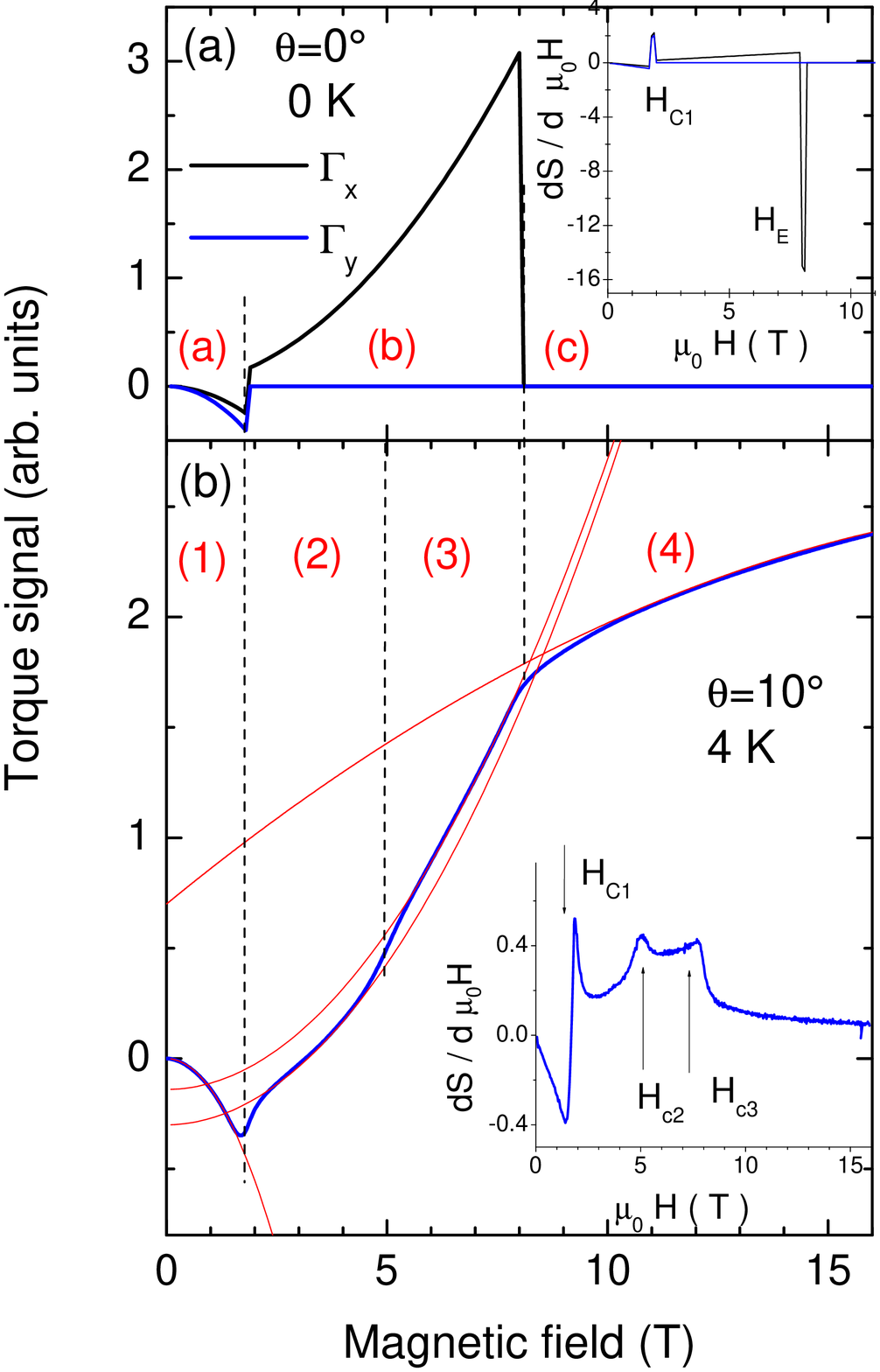}
  \caption{Magnetic field dependence of the torque on a NaNiO$_2$ crystal:(a) calculation within the easy plane model, (b) measurements and
second order polynomial fits. Inserts : derivative of the
torque.}\label{Torque}
\end{figure}
A sample with a magnetic moment $\overrightarrow{m}$ placed in a
magnetic field $\overrightarrow{H}$ will experience a torque
$\overrightarrow{\Gamma}=\overrightarrow{m}\times\overrightarrow{H}$
that can be measured using a cantilever (see figure
\ref{crystal}). One of the larger crystals was
 selected (typically 1 $mm$ x 1 $mm$ x 0.2 $mm$)
and glued on a copper-beryllium cantilever with General Electric
varnish. Care was taken to avoid contact with air and moisture.
The large face of the crystal containing the Ni planes was
parallel to the cantilever $(x,y)$ plane so that the $z$ axis
coincides with the crystal $c^{'}$ axis (see figure
\ref{crystal}). The magnetic field $H$ lies in the $yz$ plane,
making an angle $\theta$ with the $z$ axis. The cantilever could
rotate as regards the magnetic field so that $\theta$ could be
varied from 0$^{\circ}$ to 90$^{\circ}$. The capacitive technique
allows the measurement of the torque produced in the $x$
direction: $\Gamma_x=m_yH_z-m_zH_y$ where $m$ is the sample
magnetic moment. It is reduced to $\Gamma_x=\Delta\chi H^{2}
\sin(\theta-\theta_0) \cos(\theta-\theta_0)$ when the sample
magnetic moment has a linear dependence on magnetic field.
$\Delta\chi$ is the difference of the magnetic susceptibility
parallel and perpendicular to the $z$ axis; $\theta_0$ is the spin
direction in absence of torque. When the magnetization is
saturated, the torque should vanish except for $g$ factor
anisotropy effects, which should be negligible compared to
$\Delta\chi$ effects in the ordered magnetic phase.

The expected torque for $\theta=0^{\circ}$ using the easy plane
model with the small anisotropy described above is given in figure
\ref{Torque}(a), assuming that the $z$ direction coincides with
the spin direction. In the experiment, there is a 10$^{\circ}$
difference. Because of the crystal twinning, the $x$ and $y$
direction are not crystallographically well defined. This means
that the measurement gives the average of $\Gamma_x$ and
$\Gamma_y$. In these calculations, the torque has a quadratic
dependence on the magnetic field with a change of sign and
magnitude at the spin flop field ($\mu_0H_{c1}$=1.8~T) and
vanishes at either $H_{c1}$ or $H_{E}$. We expect then to observe
three different regimes labelled (a), (b) and (c) in figure
\ref{Torque}(a). The measured data are reproduced in figure
\ref{Torque}(b). Four different regimes are observed and no
saturation. Regime (1) has the expected quadratic field dependence
and can be identified as regime (a).  It disappears at $H_{c1}$ to
enter regime (2) with a change of slope, both in sign and
magnitude as expected for regime (b). This transition is widen,
probably due to thermal effects. Regime (b) is in fact divided in
two parts, (2) and (3), with similar magnetic field dependencies.
Note that the quadratic field dependence for these regimes does
not extrapolate to zero as calculated in regime (b) but rather
there is an internal torque which has different values for regime
(2) and (3). This unexpected zero field torque can exists if the
extrapolated zero field magnetization is not zero (in a canted
antiferromagnet for instance) or there exist local fields that are
not collinear with the sample magnetization (dipolar fields due to
twinning or other anisotropic fields due to crystal field effects
for instance). The presence of two regimes instead of one suggests
that there is a change in the anisotropy fields, i.e. a spin
reorientation. The higher field regime observed, regime (4), does
not saturate at all contrary to what has been calculated for
regime (c). Rather, it has the following field dependence:
$\Gamma_x=\Gamma_0+\alpha (H-H_{c4})^{2}$ with $\mu_0H_{c4}$ =
20~T. The torque starts to decrease only above $H_{c4}$, which
introduces another critical field in the magnetic phase diagram of
NaNiO$_2$. There is an internal torque in this regime as in regime
(2) and (3) with a different value. Note that these internal
torques do not exist at zero field since the corresponding
magnetic phases exist only at higher field. A domain pattern would
have developed to compensate it and have the crystal at rest.

\begin{figure}
  \includegraphics[width=8.5cm]{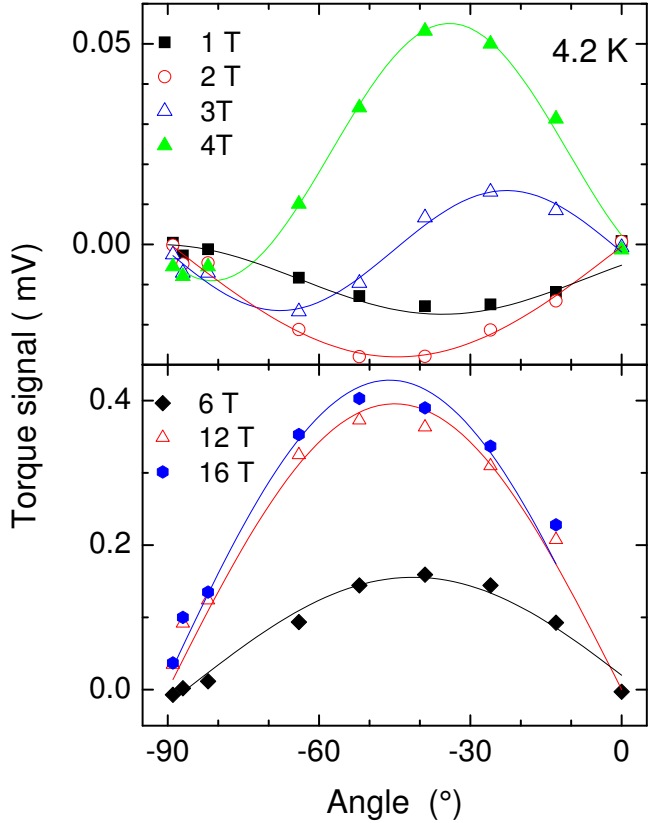}
  \caption{Angular dependence of the torque on the NaNiO$_2$ crystal at different field
  values}\label{TorqueAng}
\end{figure}
We have investigated more deeply each regime looking at their
dependence on the magnetic field orientation (see figure
\ref{TorqueAng}). The low field regime, regime (1), does not
follow at all the expected $\sin(\theta-\theta_0)
\cos(\theta-\theta_0)$ law. Rather, it behaves as $(1-\sin
3.2(\theta-105^{\circ}))$ with a periodicity of 110$^{\circ}$. The
origin of such a weird periodicity remains unclear. The presence
of an incommensurate magnetic phase is to ruled out since neutron
diffraction data at zero field are unambiguously fitted within a
commensurate magnetic structure. On the other hand, 110$^{\circ}$
is exactly the monoclinic $(a,c)$ angle of the crystal structure.
Macroscopic effects linked to twinning  are then probably
responsible but difficult to evaluate due to the low symmetry of
the system. Note that the critical field at which this regime
disappears has the following angular dependence:
$H_{c1}(\theta)=H_{c1}/\cos(\theta-\theta_{01})$ with
$\mu_0H_{c1}$=1.7~T in agreement with previous measurements
\cite{BRI05,BON66}. The free spin direction is at
$\theta_{01}=-10^{\circ}$ in  agreement with the neutron
diffraction
data \cite{DAR05}. \\

Regime (3) has the  correct angular dependence:
$\Gamma_(\theta)=\Gamma_0 \sin (\theta-\theta_0)\cos
(\theta-\theta_0)$ with $\theta_0$ close to $0^{\circ}$. Regime
(2) seems intermediate between regime (1) and (3) with a constant
evolution of the angular periodicity.

The critical fields $H_{c2}$ and $H_{c3}$ at which regime (2) and
(3) disappear follow the same angular dependence:
$H_{c}(\theta)=H_{c}/\cos((\theta-\theta_0)/2)$ with
$\mu_0H_{c.2}$=5.1~T, $\mu_0H_{c3}$=7.4~T,
$\theta_{02}=-10^{\circ}$ and $\theta_{03}=+30^{\circ}$. This
confirms the spin reorientation process from regime (2) to (3).
$H_{c3}$ is in agreement with the powder data
(\cite{BAK05,BRI07,HOL04}) while $H_{c2}$
was not detected previously.\\
Finally, regime (4) follow  the $\sin(\theta-\theta_0)
\cos(\theta-\theta_0)$ law with $\theta_0$ close to $0^{\circ}$.
We have checked that the field dependence remains the same for all
angles, with a constant value for $\mu_0H_{c4}$ around 20~T.

\begin{figure}
  \includegraphics[width=8.5cm]{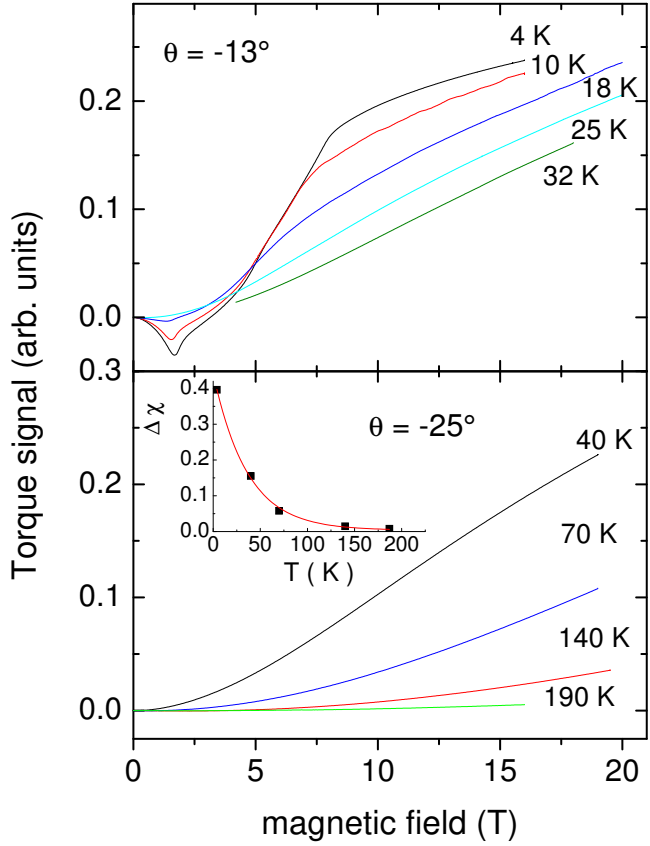}
  \caption{Magnetic field dependence of the torque at different temperatures. Insert: torque parameter $\Delta\chi$ as a function of temperature, the red line is an exponential fit.}
  \label{TorqueT}
\end{figure}

\section{Discussion}

These torque measurements on a twinned crystal reveal that the
NaNiO$_2$ system presents several spin reorientations, and very
unusual magnetic anisotropies depending on magnetic field. The
modelization is quite complex due to the low crystal symmetry of
the compound (monoclinic unit cell) and the presence of 3 types of
twins. Nevertheless, we can establish a magnetic phase diagram
using these torque measurements at different temperatures (see
figure \ref{TorqueT}) . As the ordering temperature is approached,
regime (1) disappears together while regime (2) and (3) merge into
another regime. Note that the torque diminishes as the temperature
increases but has still a non vanishing field dependence. We
checked that it behaves as $\Gamma_x=\Delta\chi H^{2}
\sin(\theta-\theta_0) \cos(\theta-\theta_0)$ and noticed that
$\Delta\chi$ decreases exponentially with an activation energy
exactly equals to the Curie Weiss temperature +36~K (see insert of
figure \ref{TorqueT}). One concludes that there is a magnetic
anisotropy induced by the exchange energy. It is responsible for
the high field torque observed at low temperature which tends to
vanish only above $H_{c4}$.

\begin{figure}
  \includegraphics[width=8.5cm]{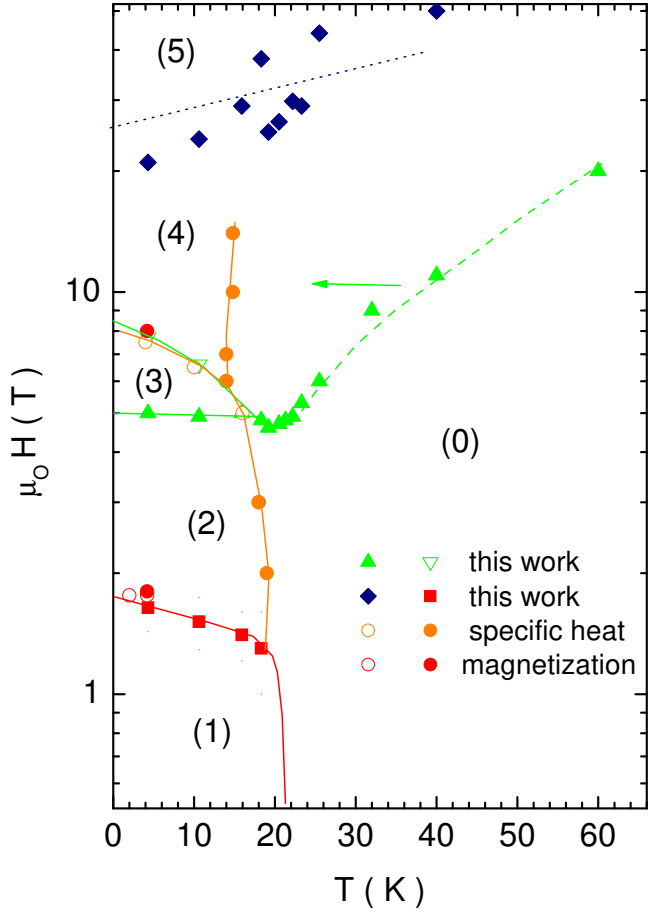}
  \caption{Magnetic phase diagram of NaNiO$_2$ from crystal
torque measurements obtained at $\theta=10^{\circ}$, heat capacity
(\cite{BAK05})and magnetization (\cite{BON66, HOL04})
measurements.}\label{Diagram}
\end{figure}

We can now establish the magnetic phase diagram. We have plotted
in figure \ref{Diagram} the critical fields obtained in this work
as well as in previous studies mainly on powder samples
\cite{BAK05, HOL04} except for the first magnetization
measurements which were taken on a twinned crystal up to 3~T
\cite{BON66}. At low temperature, 5 phases are present: phase (1)
where the spin lies at 10$^{\circ}$ from the Ni plane, phases (2)
and (3) with different spin orientations, and finally phases (4)
and (5) which are close to full saturation. Phases (1),(2)
 and (3) disappear at $T_{\rm N}= 20~K$  but a cross over line persists
 above. This phase diagram was obtained when the magnetic field was
 at 10$^{\circ}$ from the $z$ axis. All the boundary lines correspond, to a good approximation, to the phase
diagram when the magnetic field lies in the free spin direction
i.e. -10$^{\circ}$ for phase (1), +30$^{\circ}$ for phase (3),
except for the transition between phase (4)  and (0) where the
border line shifts to lower temperatures when $\theta$ is changed,
in agreement with the specific heat data on a powdered sample
\cite{BAK05} where a quasi vertical line is observed. While most
of the phase diagram was already established, we have shown, using
torque measurements on a crystal, that two additional phases are
present at low temperature, phases (3) and (5). The transition
line to phase (5), $H_{c4}(T)$, is obtained from the fit of the
torque field variations. Since it is not a direct experimental
measurement, its uncertainty is wider than for $H_{c1}$, $H_{c2}$
and $H_{c3}$. Note that $H_{c4}\simeq20~T$ corresponds to an
energy for the S=1/2 Ni$^{3+}$ spin of the order of 20~K, the same
order of magnitude as the observed anisotropy activation energy
and Curie-Weiss temperature of 36~K. As for the transition lines
$H_{c2}(T)$ and $H_{c3}(T)$, which are quite close, it is tempting
to  identify them with the two zero field modes at 6.5~$cm^{-1}$
and 9.0~$cm^{-1}$ observed in the electron spin resonance
measurements on a powdered sample at 4~K\cite{BRI07}. The
associated magnetic fields can be matched to $H_{c2}$ and $H_{c3}$
assuming a gyromagnetic factor $g\simeq$=2.6, in reasonable
agreement with the observed $g$ value at 200~K :
$g_{\parallel}$=2.03 and $g_{\perp}$=2.28. This suggests that both
$H_{c2}$ and $H_{c3} $ are associated with the same
crystallographic plane as $g_{\perp}$ (perpendicular to the
elongation axis of the oxygen octahedra). Note also that the
presence of these two modes at 6.5~$cm^{-1}$ and 9.0~$cm^{-1}$
cannot be explained in an easy plane A type antiferromagnetic
model  with only two magnetization sublattices \cite{BRI07}. The
inadequacy of this model is confirmed by our torque measurements
in regime (2) and (3) where an internal torque has to be added to
the model.

\section{Conclusion}
The magnetic study of a millimeter size NaNiO$_2$ crystal was
achieved successfully using torque measurements. We have
established the presence of 5 different ordered phases with
different spin orientations and magnetic anisotropies. The
corresponding critical magnetic fields are 1.8~T, 5.1~T, 7.4~T and
$\sim$20~T at 4~K. These reflect the presence of several
characteristic energies in the system at around 2.5~K,7.5~K, 11~K
and 30~K. A more complex behavior is observed than a simple easy
plane A type antiferromagnet  with probably different possible
magnetic ground states quite close in energy.
\\
\\
\bibliographystyle{unsrt}

\end{document}